\documentclass[a4paper,preprint,superscriptaddress,preprintnumbers,nofootinbib]{revtex4}

\usepackage[dvipdf,dvips,dvipdfmx]{graphicx}
\usepackage{amsmath}
\usepackage{amssymb}
\usepackage{float}
\usepackage{wrapfig}
\usepackage{bm}
\usepackage{color}

\def\lsim{\mathrel{\mathpalette\@versim<}}
\def\gsim{\mathrel{\mathpalette\@versim>}}

\newcommand{\al}[1]{\begin{align}#1\end{align}}
\newcommand{\bp}{\begin{pmatrix}}
\newcommand{\ep}{\end{pmatrix}}
\newcommand{\nn}{\nonumber\\}
\newcommand{\paren}[1]{\left(#1\right)}
\newcommand{\sqbr}[1]{\left[#1\right]}

\newcommand{\vev}[1]{\left\langle#1\right\rangle}

\newcommand{\p}{\partial}

\newcommand{\red}{\textcolor{black}}

\newcommand{\GeV}{\,\text{GeV}}

\newcommand{\eV}{\ensuremath{\,\text{eV} }}

\newcommand{\bs}[1]{\boldsymbol}

\newcommand{\pmat}[1]{\begin{pmatrix}#1\end{pmatrix}}

\newcommand{\fn}[1]{\!\left(#1\right)}

\newcommand{\bra}{\langle}
\newcommand{\ket}{\rangle}
\newcommand{\Lag}{\mathcal L}

\begin{document}

\title{Baryogenesis in false vacuum}

\author{Yuta \surname{Hamada}}
\affiliation{KEK Theory Center, IPNS, KEK, Tsukuba, Ibaraki 305-0801, Japan}

\author{Masatoshi \surname{Yamada}}
\affiliation{Institute for Theoretical Physics, Kanazawa University, Kanazawa 920-1192, Japan}

\preprint{
KEK-TH-1907
}
\preprint{
KANAZAWA-16-7
}
\begin{abstract}
The null result in the LHC may indicate that the standard model is not drastically modified up to very high scales such as the GUT/string scale.
Having this in the mind, we suggest a novel leptogenesis scenario realized in the false vacuum of the Higgs field.
If the Higgs field develops a large vacuum expectation value in the early universe, \red{a} lepton number violating process is enhanced, which we use for baryogenesis.
To demonstrate the scenario, several models are discussed.
For example, we show that the observed baryon asymmetry is successfully generated in the standard model with higher-dimensional operators.
\end{abstract}
\maketitle

\section{Introduction}
Although the standard model (SM) is complete after the discovery of the Higgs boson at the Large Hadron Collider~\cite{Aad:2012tfa,Chatrchyan:2012xdj}, there are still mysteries in elementary particle physics, such as the finite neutrino mass and dark matter.
Besides these the baryon asymmetry in the universe (BAU) is also one of the unsolved problems.
That is, how \red{has} baryogenesis been realized in the evolution of the universe?
The latest cosmological result from the Planck observations~\cite{Ade:2015xua} tells us that the BAU is 
\al{\label{bnas}
\frac{n_B}{s}=
(8.67\pm 0.05)\times 10^{-11},
}
where $n_B$ is the baryon number density and $s$ is the entropy density.

In order to theoretically explain the BAU within elementary particle physics, the Sakharov conditions~\cite{Sakharov:1967dj} have to be satisfied: there exists a process violating the baryon number conservation; $C$ and $CP$ invariances are violated; the system leaves its equilibrium state.
The SM does not accommodate the departure from equilibrium.
Although the baryon number is violated  through the sphaleron process and $CP$ symmetry is violated in the weak interaction, it is not enough to reproduce the BAU.
Therefore the SM cannot satisfy these conditions and must be extended.

Some baryogenesis mechanisms satisfying the Sakharov conditions have been suggested, e.g. the grand unified theory~\cite{Yoshimura:1978ex} and the Affleck--Dine mechanism~\cite{Affleck:1984fy}. 
Leptogenesis is also one of the well-known mechanisms for baryogenesis ~\cite{Fukugita:1986hr} (see also the reviews~\cite{Buchmuller:2005eh,Fong:2013wr}) where we use the fact that through the sphaleron process~\cite{Dashen:1974ck,Manton:1983nd,Klinkhamer:1984di,Kuzmin:1985mm}, the difference $B-L$ between the baryon number $B$ and the lepton number $L$ is conserved whereas their sum $B+L$ is not.
The baryon number density in thermal equilibrium is provided by the $B-L$ number density via the sphaleron process:
\al{\label{sphaleronbl}
n_B=\frac{8N_F+4N_S}{22N_F+13N_S}n_{B-L},
}
where $N_F$ is generation of quarks and leptons and $N_S$ is that of scalar doublets.
For instance, in the case of the SM where $N_F=3$ and $N_S=1$, the factor in the right-hand side is $28/79$.
Through the decay of the heavy particle, the lepton number is generated, and then its number density changes to the $B-L$ number density $n_{B-L}$, whose process is described by the coupled Boltzmann equations for these number densities.

In this paper, we study leptogenesis realized in the false vacuum of the Higgs field, in which the Higgs gains a vacuum expectation value far above the electroweak scale.
The mass of the particles coupled to the Higgs field becomes super-massive, and the left-handed neutrinos can become heavier than the charged leptons and $W$ boson in the presence of higher-dimensional lepton number violating operators. 
The decay of the left-handed neutrinos creates an $L$ asymmetry.
Shortly after, the phase transition of the Higgs takes place and the Higgs moves from the false vacuum to the true electroweak one (where thermal effects restore the electroweak symmetry allowing the sphalerons to reprocess the $L$ into a $B$ asymmetry).
We consider the situation where the right-handed neutrino masses are large compared with the reheating temperature. 
Therefore our scenario gives an alternative scenario for baryogenesis.


To demonstrate this scenario, two models are investigated. 
We first consider a minimal model depending on the SM with a higher-dimensional operator,
\al{\label{dimfiveop}
\Delta \Lag_5 =\frac{\lambda_{ij}}{\Lambda} H H \bar L^c_j L_i,
}
where $L_i$ is the lepton doublet, $\Lambda$ is a cutoff scale,\footnote{
In Refs.~\cite{Aoki:1997vb,Hamada:2015xva}, the operator \eqref{dimfiveop} is used to realize leptogenesis as well as the $CP$ violating operator $\bar{L}_i\gamma^\mu L_j\bar{L}_i\gamma_\mu L_j$.
These operators are naturally generated in the low energy effective theories of various seesaw models.
} and the Higgs doublet is defined as
\al{\label{higgsdoublet}
H =\frac{1}{\sqrt{2}} 
\pmat{
\chi_1 +i \chi_2\\
h+ i \chi_3
}.
}
Such an operator is typically generated in the type I seesaw model by integrating out the right-hand neutrino.
This effective interaction breaks the lepton number conservation and thus is used as the source of the lepton asymmetry.
In particular, we consider the decays of the left-hand neutrino, given by the modes $\nu \to \ell^- W^+,\, \ell^+ H^-, \, \ell^- H^+$.
Note that in the broken phase ${\vev H}\neq 0$, the operator \eqref{dimfiveop} turns into a neutrino mass term,
\al{\label{Eq:neutrino mass}
\frac{\vev{H}^2}{\Lambda}(\bar \nu^c\nu + \bar \nu \nu^c)
=
\frac{\vev{h}^2}{2\Lambda}(\bar \nu^c\nu + \bar \nu \nu^c)
,
}
where we have assumed that the coupling constant $\lambda_{ij}$ is of order one since neutrino can have a finite mass $m_\nu\sim 0.1$~eV.
Then the cutoff scale $\Lambda$ is estimated as
\al{\label{cutoff}
\Lambda\simeq6.0\times10^{14}\GeV\paren{0.1\eV\over m_\nu}.
}
Here, leptogenesis takes place in the false vacuum where the neutrino mass $\vev{h}^2/\Lambda$ becomes larger than the charged lepton and the $W$ boson ones.
As will be seen in the next section, in such a minimal model, the baryon asymmetry produced by this process actually is not adequate for the observed value \eqref{bnas}.

Next, we consider an extended system in which the new higher-dimensional operators are added.
In this case, we will see that the lepton asymmetry is caused by the neutrino and it is possible to explain the observation.

We have to see whether or not the phase transition of the Higgs field from the false vacuum to the electroweak one occurs after the lepton asymmetry is produced.
To this end, we investigate the thermal history of the Higgs potential.
Including a new singlet-scalar field coupled to the SM Higgs field, there exists a certain parameter space where the phase transition appropriately takes place.

We organize this paper as follows:
in the next section, we present the formulation of the Boltzmann equations in order to calculate the baryon asymmetry.
Numerically solving them, we investigate the produced baryon asymmetry for two cases explained above.
Section~\ref{Thermal history} is devoted to an investigation of the thermal history of the Higgs potential.
We summarize and discuss our study and the results obtained, and we comment on the possibility of the high scale electroweak baryogenesis in the section~\ref{Summary and discussion}.
In the appendix~\ref{The effective potential at finite temperature}, the thermal effects on the Higgs potential and their formulations are shown.

\section{Mechanism and Boltzmann equations}\label{Mechanism and Boltzmann equations}
First, we consider a situation where the decay of the left-handed neutrino produces the baryon asymmetry.
In this section, we present the Boltzmann equations and quantitatively evaluate the baryon asymmetry by numerically solving them.
We evaluate the baryon asymmetry produced by the left-handed neutrino decay; however, we see that not enough baryon asymmetry is produced.
To ameliorate the situation, next we add the new higher-dimensional operators.
We demonstrate that, in this case, the decay of the neutrino can reproduce the observed amount of asymmetry.
%

\subsection{The derivation of Boltzmann equations}
In this subsection, to calculate the asymmetry of the universe, we follow Refs.~\cite{Kolb:1979qa,Kolb:1990vq,Fong:2013wr} and derive the Boltzmann equations for the general case of leptogenesis.
The change of the number density of a heavy particle is governed by
\al{
\label{Eq:Boltzmann0}
\dot{n}_X+3Hn_X
&=
\int d\Pi_X \, d\Pi_1 \,d\Pi_2 \,(2\pi)^4\delta^{(4)}(p_X-p_1-p_2) 
\nn
&\times
\paren{
-f(p_X) |\mathcal{M}(X\rightarrow 1 2)|^2 + f(p_1)f(p_2)  |\mathcal{M}(12\rightarrow X)|^2
}
\nn
&+
\int d\Pi_X \, d\Pi_Y \, d\Pi_1 \, d\Pi_2 \,\cdots d\Pi_N \,(2\pi)^4\delta^{(4)}(p_X+p_Y-p_1-p_2-\cdots-p_N)
\nn
&\times
 \paren{
-f(p_X)f(p_Y) |\mathcal{M}(XY\rightarrow12\cdots N)|^2 + f(p_1)\cdots f(p_N)  |\mathcal{M}(12\cdots N\rightarrow XY)|^2
},
}
where $X$ and $Y$ represent the heavy particles
; the numbers $1\cdots N$ denote lighter particles; the dot on $n_X$ in the left-hand side denotes the time derivative; we have neglected the effects of the Pauli blocking and stimulated emission; $d\Pi_i=d^3p_i/(2\pi)^32E_i$ is the phase space integral; $H={\dot R}/R$ is the Hubble parameter given by the scale factor $R$ which is governed by the Friedmann equation. Here $f$ is the distribution function, approximately given by the Maxwell-Boltzmann distribution.

The first and second terms of the right-hand side in Eq.~\eqref{Eq:Boltzmann0} correspond to the decay and annihilation of heavy particle, respectively.
Let us rewrite the first term by using the definition of the decay rate,
\al{
\Gamma_X
\equiv
{1\over2E_X}
\int d\Pi_1 \,d\Pi_2 \,(2\pi)^4\delta^{(4)}(p_X-p_1-p_2) 
|\mathcal{M}(X\rightarrow 1 2)|^2.
}
We use the fact that the kinetic equilibrium allows us to make the replacement,\footnote{
Here we neglect the chemical potential of particle 1 and 2 as the effect is subleading.
}
\al{
f(p_1)f(p_2)
=
f^\text{EQ}(p_1)f^\text{EQ}(p_2)
=
f^\text{EQ}(p_X).
}
Furthermore, at leading order, $|\mathcal{M}(X\rightarrow12)|^2=|\mathcal{M}(12\rightarrow X)|^2$.
Hence, we find that the first term in the right-hand side becomes
\al{
\paren{
-n_X  + n_X^\text{EQ}
}
\Gamma_X.
\label{part 1 of bolzmann}
}
The second term in Eq.~\eqref{Eq:Boltzmann0} can be written in terms of the thermal average cross section of the pair annihilation $\vev{\sigma_\text{ann} v}:$
\al{
\vev{\sigma_\text{ann} v}
=
{\int d\Pi_X \, d\Pi_Y \, d\Pi_1 \cdots d\Pi_N \,(2\pi)^4 \delta^{(4)}\fn{p_X+p_Y-p_1-\cdots-p_N}
|\mathcal{M}(XY\rightarrow1\cdots N)|^2
\over
\int d\Pi_X \, d\Pi_Y \, (2E_X) (2E_Y) f\fn{p_X} f\fn{p_Y}
}.
}
We assume that $f\fn{p_i}\propto f^\text{EQ}\fn{p_i}$ thanks to the kinetic equilibrium, so that the second term in Eq.~\eqref{Eq:Boltzmann0}  becomes
\al{
\vev{\sigma_\text{ann} v}
 \paren{
-n_X^2  + \paren{n_X^\text{EQ}}^2
}.
}
To summarize, the Boltzmann equation of $n_X$ is given by
\al{
\label{Eq:Boltzmann_X}
\dot{n}_X+3Hn_X
&=
\paren{
-n_X  + n_X^\text{EQ}
}
\Gamma_X
+\vev{\sigma_\text{ann} v}
 \paren{
-n_X^2  + \paren{n_X^\text{EQ}}^2
}.
}

In a similar manner, we can write the Boltzmann equation governing the lepton number density:
\al{
\dot{n}_l+3Hn_l
&=
\int d\Pi_X \, d\Pi_l \,d\Pi_{W} \,\delta^{(4)}\fn{p_X-p_l-p_{W}}
\nn
&\quad \times
\epsilon \paren{
-f\fn{p_X} |\mathcal{M}\fn{X\rightarrow l W}|^2 + f\fn{p_l}f\fn{p_{W}}  |\mathcal{M}\fn{l W\rightarrow X}|^2
}
\nn
&\qquad +2
\int d\Pi_1 \, d\Pi_2 \, d\Pi_3 \, d\Pi_4 \,\delta^{(4)}\fn{p_1+p_2-p_3-p_4}
\nn
&\quad \qquad \times
 \paren{
-f\fn{p_1}f\fn{p_2} |\mathcal{M}\fn{l_1 l_2\rightarrow l_3 l_4}|^2 + f\fn{p_3}f\fn{p_4}  |\mathcal{M}\fn{l_3 l_4\rightarrow l_1 l_2}|^2
},
}
where the first and second terms in the right-hand side describe the decay of the heavy particle and annihilation of the leptons, respectively; $W$ is a particle without the lepton number; $l_i$ is a particle having the lepton number symmetry; the process $X\leftrightarrow l W$ breaks the lepton number.  
Furthermore, we rewrite this equation as one for $B-L$ asymmetry, which is given by
\al{
\label{Eq:Boltzmann_B-L}
\dot{n}_{B-L}+3Hn_{B-L}
&=
-
\epsilon
\paren{
n_X  - n_X^\text{EQ}
}
\Gamma_X \,\text{Br}
-n_{B-L} \Gamma_X \,\text{Br} \,{n_X^\text{EQ}\over n_\gamma}
-2 n_{B-L}n_l \vev{\sigma_L v},
}
where $\epsilon$ is the parameter which denotes the $CP$ asymmetry; Br is the branching ratio of $X\to l W$; $n_\gamma$ is the number density of photon; and $\vev{\sigma_L v}$ is the thermally-averaged scattering cross section which does not conserve the lepton number.

It is convenient to introduce $N_i\equiv n_i/n_\gamma$ because this quantity is conserved under the cosmic expansion. 
We also introduce $z\equiv M_X/T$ as a variable.
Using these variables, let us now rewrite the Boltzmann equations.
For instance, the left-hand side becomes
\al{
\dot{n}_X+3Hn_X
=
n_\gamma
\dot{N}_X
=
n_\gamma H z
{d\over dz}
N_X,
}
where, in the second equality, we have used
\al{
{dT\over dt}
=
-3H
{n_\gamma\over dn_\gamma/dT}
=
-H T.
}
The right-hand side is
\al{
&-\Gamma_X(z)
\paren{
n_X-n_X^\text{EQ}
}
-\vev{\sigma_\text{ann}v}\paren{n_X^2-\paren{n_X^\text{EQ}}^2}
\nn
&=
-n_\gamma Hz
\paren{\Gamma_X(z)\over H(z)z}
(N_X-N_X^\text{EQ})
-n_\gamma Hz
\paren{\vev{\sigma_\text{ann}v}n_\gamma\over H(z)z}
\paren{N_X^2-\paren{N_X^\text{EQ}}^2}
.
}

In terms of $N_i$ and $z$, we can write the set of the Boltzmann equations as follows:
\al{
{d\over dz}N_X
&=
-\paren{\Gamma_X(z)\over H(z) z}
(N_X-N_X^\text{EQ})
-
\paren{\vev{\sigma_\text{ann}v}n_\gamma \over H(z)z}
\paren{N_X^2-\paren{N_X^\text{EQ}}^2}
\label{boltzmann1}
,
\\
{d\over dz} N_{B-L}
&=
-\paren{\epsilon\Gamma_X(z)\text{Br}\over H(z) z}
(N_X-N_X^\text{EQ})
-N_{B-L}\paren{\Gamma_X(z)\text{Br}\over H(z) z}{N_X^\text{EQ}}
\nn
&\phantom{=}
\quad -\paren{\vev{\sigma_L v}n_\gamma\over H(z)z}2N_{B-L} N_l
,
\\
H^2\fn{z}
&=
{\pi^2 g_*(z)\over90}
{z^4 M_X^4\over M_P^2},
\\
N_X^\text{EQ}
&
=\frac{g}{4\zeta\fn{3}}z^2K_2\fn{z}
,
\\
n_\gamma
&=
{2\zeta(3)\over\pi^2}T^3
={2\zeta(3)\over\pi^2} z^{-3}M_X
,
\label{photondensity}
}
where $\zeta\fn{3}\approx 1.20205$ is the Riemann zeta function of 3; $M_P=\sqrt{\hbar c/8\pi G}=2.435\times 10^{18}$ GeV is the reduced Planck scale; $K_2$ is for the modified Bessel functions of the second kind; $g_*\fn{z}$ is the total number of effectively massless degrees of freedom; and $g$ is the internal degrees of freedom of the heavy particle.
We neglect the $z$ dependence of $g_*\fn{z}$ and use $g_*=106.75$.

Simultaneously solving the Boltzmann equations, we can evaluate the value of the lepton asymmetry due to the decay of the left-handed neutrino which is identified with the heavy particle $X$.
In order to perform numerical calculations, we have to specify $\Gamma_X, \text{Br}, \epsilon, \vev{\sigma_\text{ann} v}$ and $\vev{\sigma_L v}$.
In the next subsection, we give these variables for the minimal model.

\subsection{Minimal model case}
We evaluate the baryon asymmetry in the minimal model whose Lagrangian is given as
\al{
\Lag=\Lag_\text{SM}+\Delta \Lag_5,
}
where $\Lag_\text{SM}$ is the Lagrangian of the SM and $\Delta \Lag_5$ is the higher-dimensional operator given in Eq.~\eqref{dimfiveop}.
The lepton number is produced by the decay of the left-handed neutrino.
We now show the variables given in the Boltzmann equations in order.

The masses of the left-handed neutrino, the Higgs boson and the $W$ boson in the broken phase $\vev{H} \neq 0$ are given as
\al{
m_\nu
&=
{\vev{h}^2\over \Lambda},&
M_H&=\sqrt{2\lambda}\vev{h},&
M_W&=
{1\over2}g_2\vev{h},&
}
respectively, where $\lambda$ is the quartic coupling constant of the Higgs field, and $g_2$ is the $\text{SU}(2)_L$ gauge coupling constant. 

The decay rate of the left-handed neutrino, $\nu_i$, and the branching ratio to the longitudinal gauge boson are  calculated as\footnote{
Here we assume that $\lambda_{ij}$ is the order of $1$ quantity, as in Eq.~\eqref{Eq:neutrino mass}.
}
\al{&\label{Eq:decay rate}
\Gamma_X\fn{z}
=
\left\bra \frac{1}{\gamma} \right\ket \Gamma_X|_{z=\infty}
\simeq\left\bra \dfrac{1}{\gamma}\right\ket  \dfrac{m_\nu}{8\pi}
\paren{
\paren{\lambda_{ii}{\vev{h}\over\sqrt{2}\Lambda}}^2+g_2^2+\, y_\tau^2
},
&
\text{Br}\simeq
{y_\tau^2\over\paren{{\vev{h}\over\sqrt{2}\Lambda}}^2+g_2^2}
,
}
where $\vev{1/\gamma}=K_1(z)/K_2(z)$ in the thermal bath, $K_1$ is for the modified Bessel functions of first kind, $\lambda_{ii}$ is diagonalized by rotating the lepton field, and $y_\tau$ is the tau Yukawa coupling. We note that the branching ratio to the transverse gauge boson is important.
This is because, in order to pick up the imaginary part of the amplitude, one needs to use the lepton Yukawa coupling rather than the $SU(2)$ gauge coupling.\footnote{
The coupling $\lambda_{ij}$ is not helpful \red{in obtaining} the imaginary part because this coupling becomes real by rotating the lepton field.
}

\begin{figure}
\begin{center}
\hfill
\includegraphics[width=.7\textwidth]{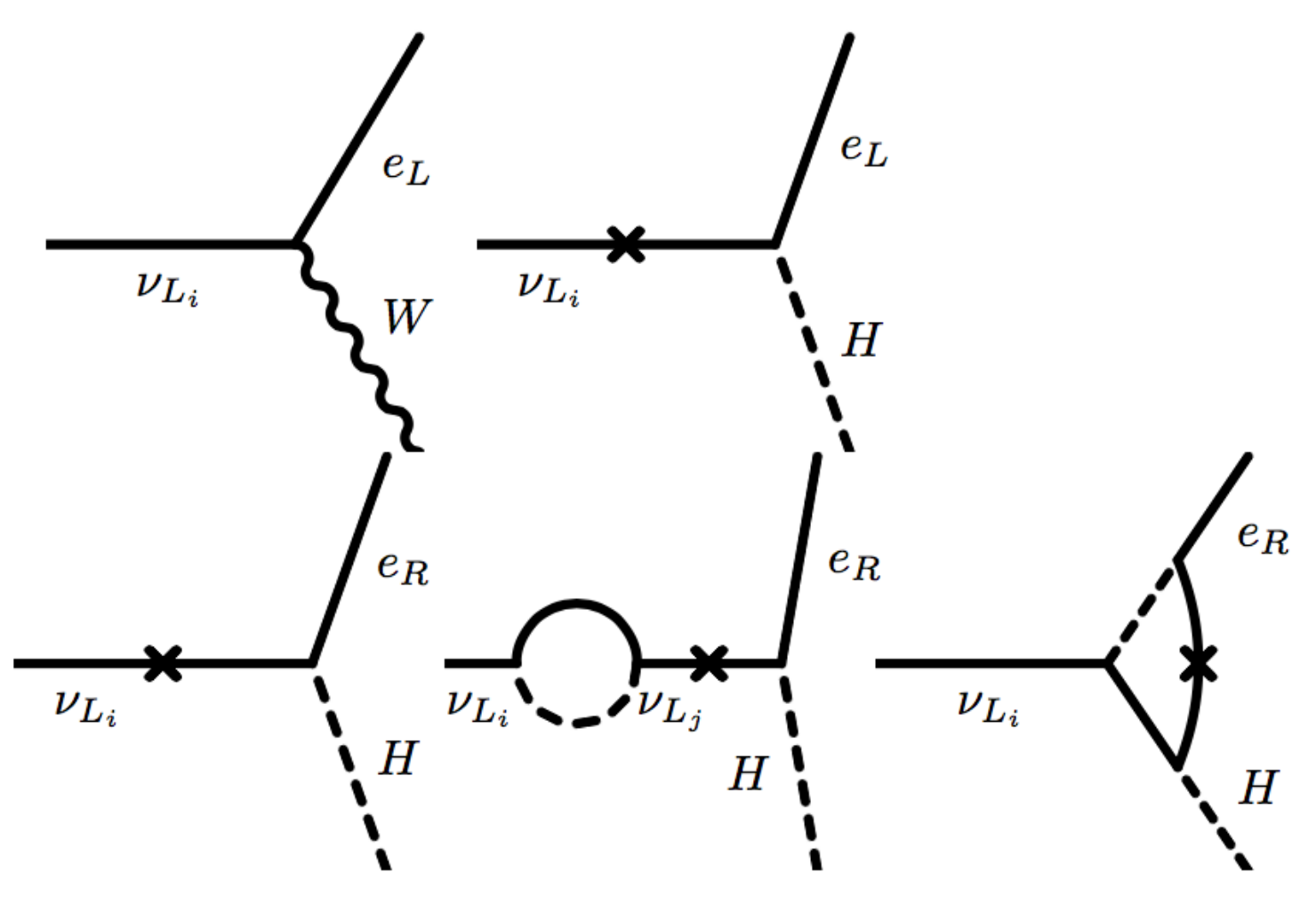}
\hfill\mbox{}
\end{center}
\caption{
The first two diagrams are the main decay modes of the neutrino where the second one comes from the vertex $\vev{H} H \bar L^c_j L_i$.
The last three diagrams contribute to the asymmetry by the decay of the left-hand neutrino.
The complex phase appears only if the mass of neutrino is larger than that of $W$ boson.
Majorana mass term for the left handed neutrino explicitly breaks the lepton number conservation, which is represented by cross symbol in the diagram.
}
\label{Fig:nuL_decay}
\end{figure}
The $CP$ asymmetry $\epsilon$ comes from the interference between the tree and the loop diagrams corresponding to the last three diagrams in Fig.~\ref{Fig:nuL_decay}, 
whose order is given by
\al{\label{Eq:epsilon}
\epsilon_i
&\simeq
\dfrac{1}{8\pi}
\sum_{j}
{\mathrm{Im}\sqbr{\paren{YY^\dagger}_{ij}^2}\over y_\tau^2},
}
%
%
where $Y$ is the charged lepton Yukawa matrix.
Note that the imaginary part appears only if $m_\nu > M_W+M_\tau$,\footnote{
Even if $m_\nu < M_W+M_\tau$, the imaginary part appears in higher order.
However, it is too small to obtain sufficient baryon asymmetry.
}
which yields
\al{
\vev{h}
>
{g_2\over2}\Lambda
\simeq
1.5\times10^{14}\GeV
\paren{\Lambda\over 6\times10^{14}\GeV}
.
}
Here $M_\tau$ is the mass of the tau lepton.

Let us estimate the imaginary parts of the Yukawa coupling constants in Eq.~\eqref{Eq:epsilon}.
The numerator of Eq.~\eqref{Eq:epsilon} is related to the Jarlskog invariant in the lepton sector~\cite{Olive:2016xmw}, and the order is estimated as~\cite{Olive:2016xmw}
\al{\label{Eq:epsilon estimation}
\epsilon_i
\sim {1\over8\pi} y_\tau^2\paren{ 3\times10^{-2}\sin\delta}
\simeq
1.2\times10^{-7}\paren{y_\tau\over10^{-2}}^2 \paren{\sin\delta\over1},
}
where $\delta$ is the Dirac CP phase of the neutrino sector.

We note that, by using the renormalization group equations, we obtain the values of the coupling constants at the high scale:\footnote{See e.g. Ref.~\cite{Hamada:2013mya}.}
\al{
g_2&\simeq\red{0.5}, &
y_\tau&\simeq1\times10^{-2},&
\alpha_2&\simeq \frac{g_2^2}{4\pi} ={1\over50}, &
}

\subsubsection*{Numerical result in minimal model}
The Planck observation~\cite{Ade:2015xua} tells us 
\al{
N_{B,\text{obs}}\simeq
6.1\times10^{-10}
\times{2387\over86}
=1.7\times10^{-8}
,
}
where the factor $2387/86$ is the photon production factor.\footnote{\red{
This factor comes from the ratio $g_{*s}(T_B)/g_{*s}(\text{today})$, where $g_{*s}$ is the effective degrees of freedom for the entropy, and $T_B$ is the temperature of the baryogenesis. Here $g_{*s}$ is given by
\al{&
g_{*s}(\text{today})={43\over11},
&&
g_{*s}(T_B)={217\over2}.
} 
The photon and the left handed neutrinos contribute $g_{*s}(\text{today})$, and we assume that the SM particles and one generation right handed neutrino contributes $g_{*s}(T_B)$.
Even if the right handed neutrino contribution is absent, this factor rarely changes.
}}
If this value comes from the sphaleron effect, we should have
\al{
N_{B-L,\text{obs}}\simeq
6.1\times10^{-10}
\times{2387\over86}
\times{79\over28}
=
4.8\times10^{-8}.
\label{NBLobserved}
}
Therefore, we numerically solve the Boltzmann equations given in Eqs.~\eqref{boltzmann1}--\eqref{photondensity} and investigate whether or not the appropriate parameter space which satisfies the value \eqref{NBLobserved} exists.

\begin{figure}
\begin{center}
\hfill
\includegraphics[width=.45\textwidth]{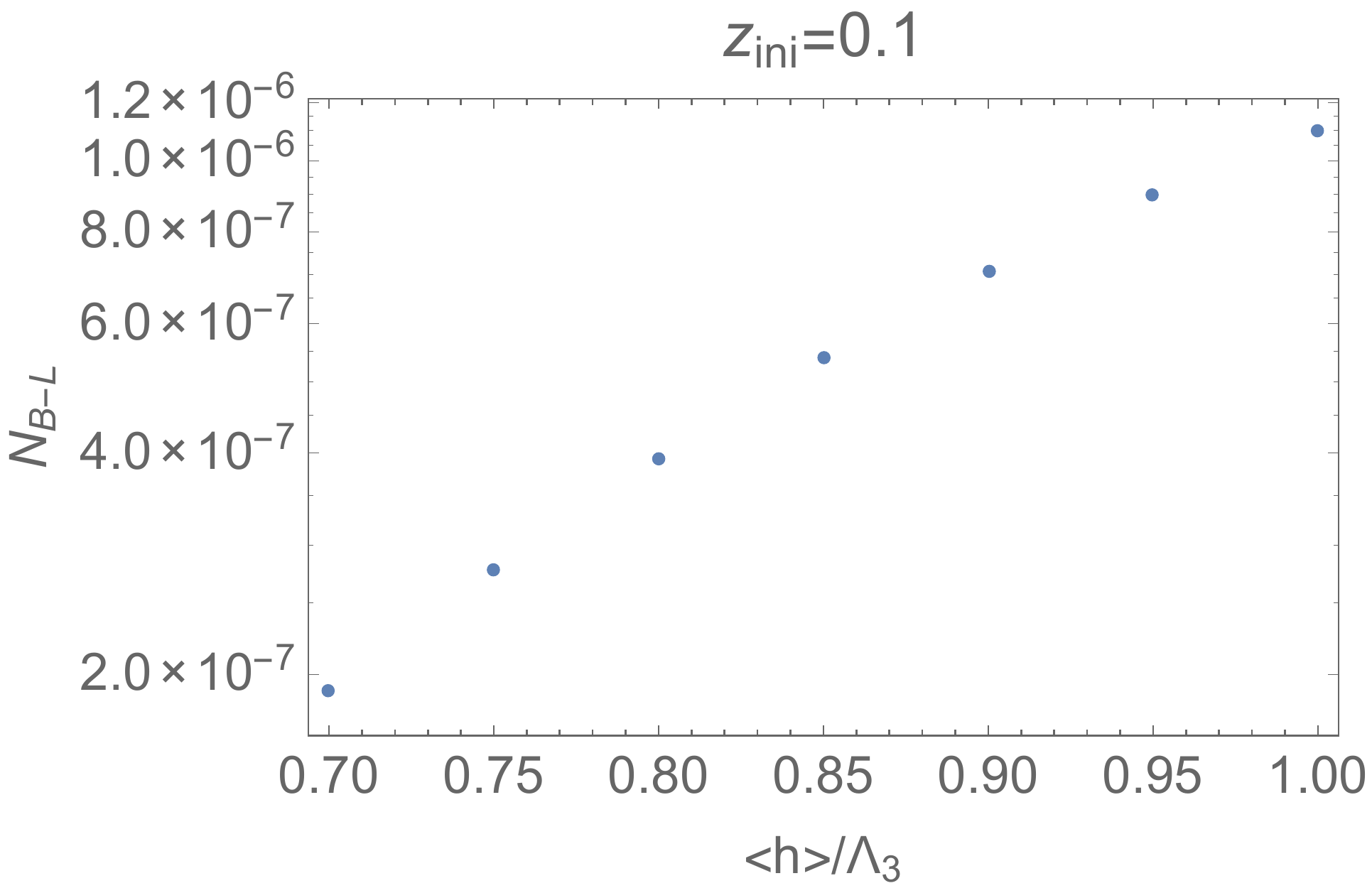}
\hfill
\includegraphics[width=.45\textwidth]{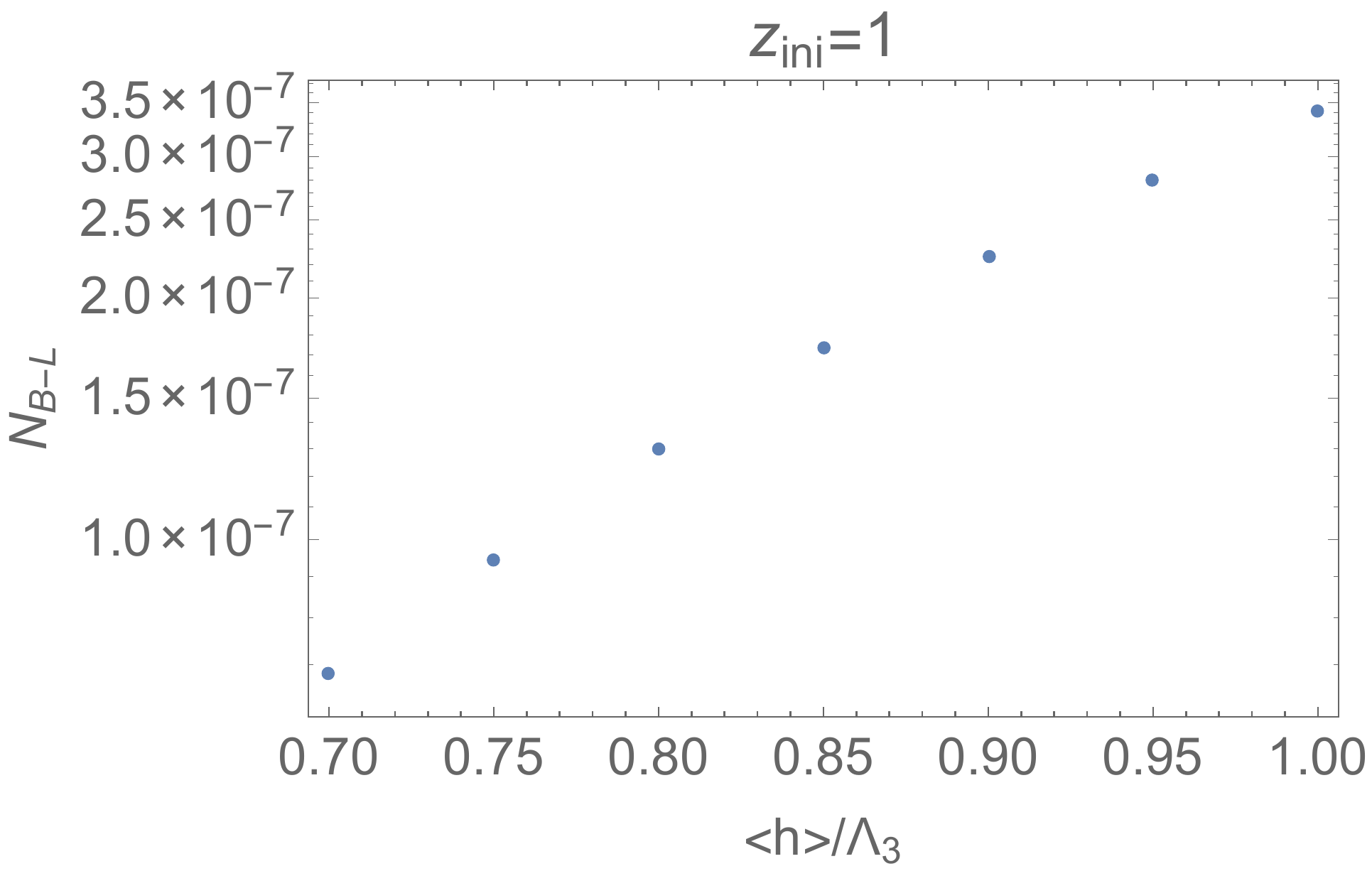}
\hfill\mbox{}
\hfill
\includegraphics[width=.45\textwidth]{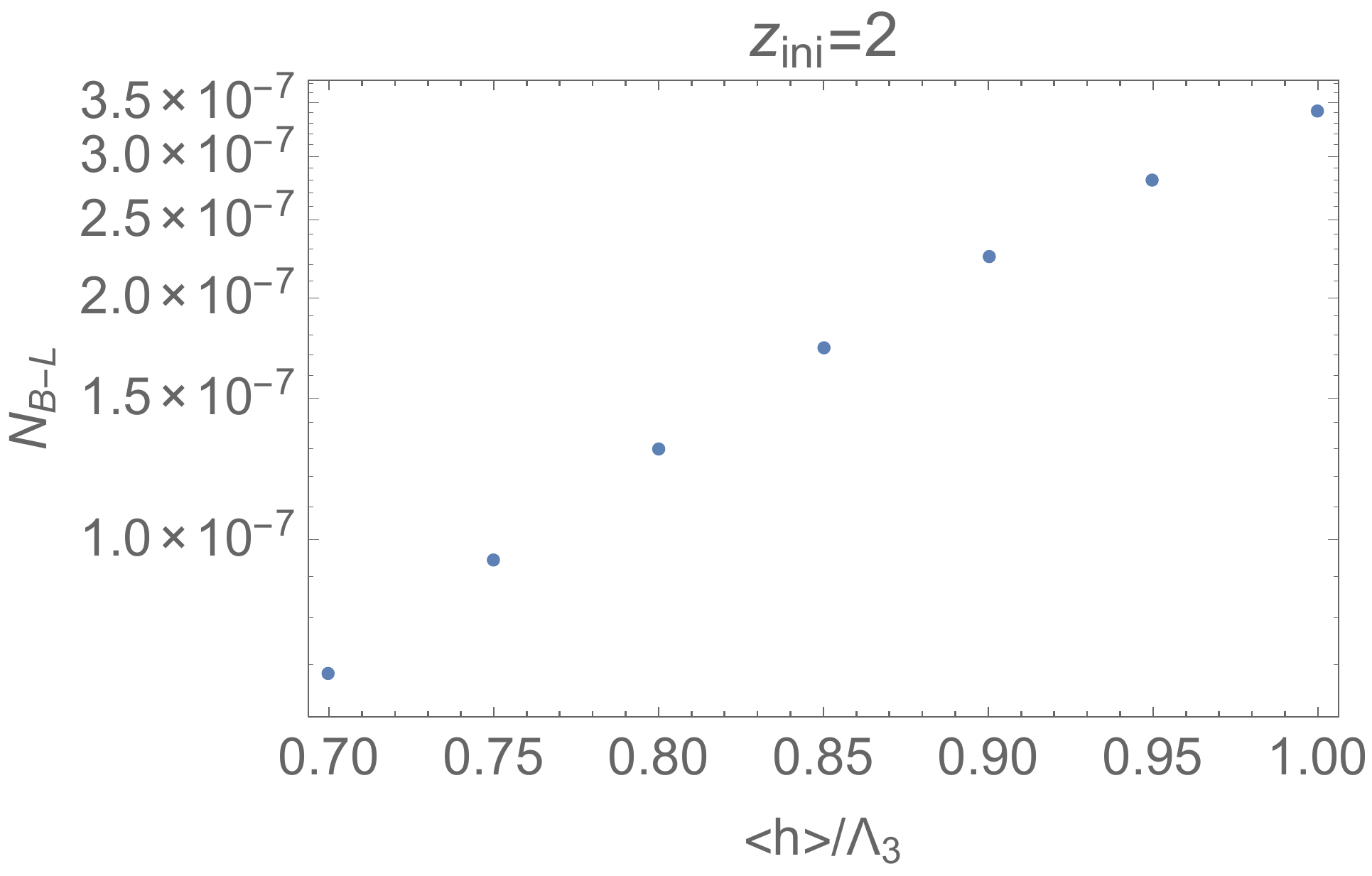}
\hfill
\includegraphics[width=.45\textwidth]{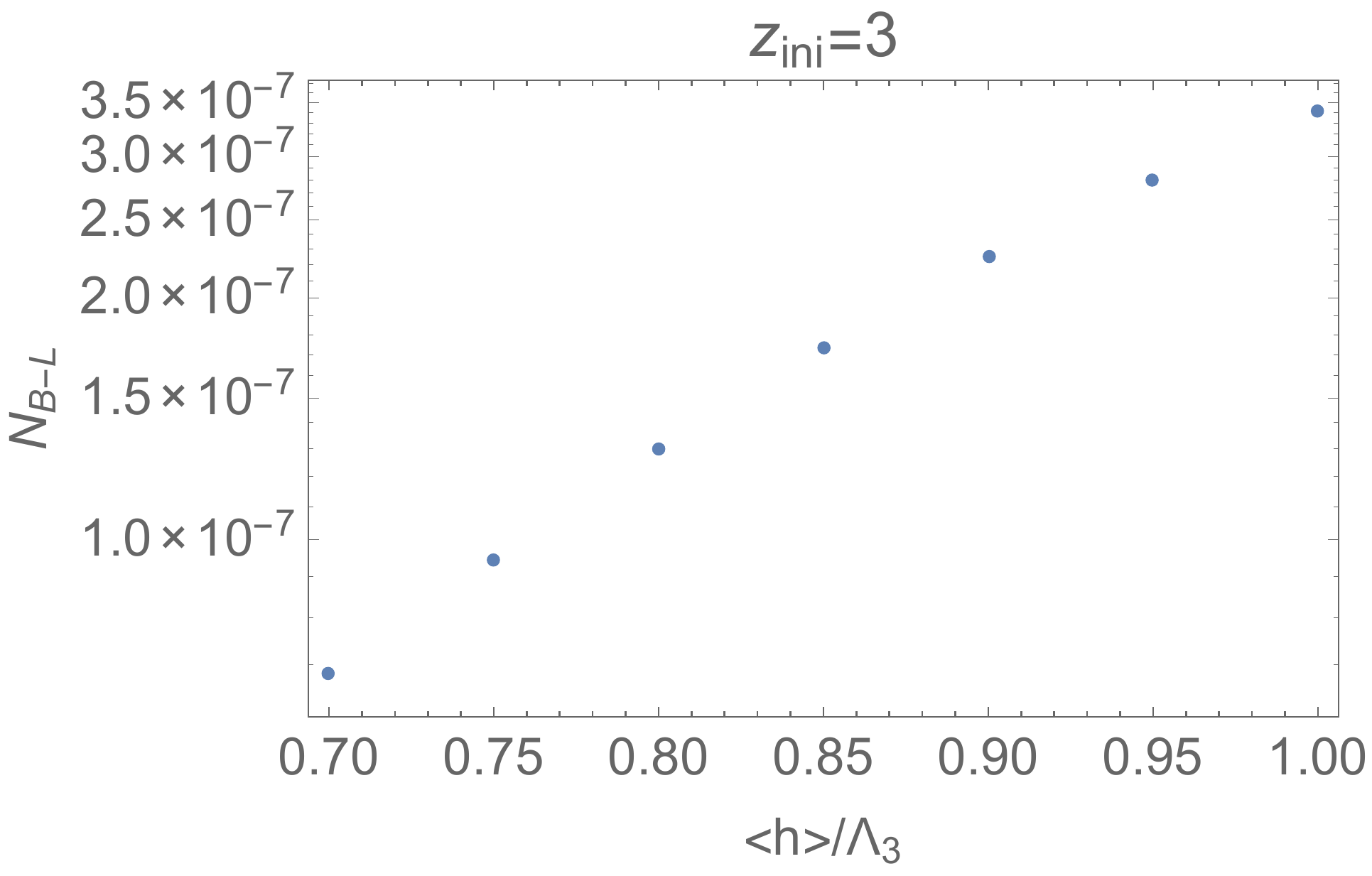}
\hfill\mbox{}
\end{center}
\caption{
$N_{B-L}$ created by the decay of neutrinos in the presence of the higher dimensional operators.
The parameters are taken as in Eq.~\eqref{Eq:parameter set}.
}
\label{Fig:charged_Higgs}
\end{figure}


%
Unfortunately, we can easily see that the baryon asymmetry cannot be reproduced in this framework.
We obtain
\al{\label{Eq:minimal model analytic}
\epsilon_i \text{Br}
\sim 
1.2\times10^{-7}{y_\tau^2\over\paren{{\vev{h}\over\sqrt{2}\Lambda}}^2+g_2^2}\paren{y_\tau\over10^{-2}}^2 \paren{\sin\delta\over1}
\lesssim
2\times10^{-10},
}
by combining Eqs.~\eqref{Eq:decay rate} and \eqref{Eq:epsilon estimation}, and hence the resultant baryon asymmetry is too small to explain the current data.
This indicates the necessity of an extension of the model.
In the next subsection, we present the possible extension to realize the observed baryon asymmetry.

\subsection{Extended model}
A way to improve the situation is to add new operator.
The smallness of the charged lepton Yukawa coupling results in the small baryon asymmetry.
Therefore, if this coupling is modified in the false vacuum, the situation changes.
Let us assume the existence of the higher-dimensional operator which contributes as the Yukawa coupling in the vacuum where Higgs takes the large VEV:
\al{
y_{2,ij} {H^\dagger H\over \Lambda_2^2} \bar{E}_i H L_j+h.c.
}
Similar, we consider the operator which gives the correction to Majorana neutrino masses:
\al{
y_{3,ij}{H^\dagger H\over \Lambda_3^3}H H \bar L^c_j L_i+h.c.
}
In general, $\Lambda_3$ can be different from $\Lambda_1$ in Eq.~\eqref{dimfiveop}.
This structure may occur when we consider the right-handed neutrino model as a UV completion for example,\footnote{This is just a possibility of the UV completion, and we do not insist on this model in the following discussion.} where the action is
\al{
M_{N,ij}\bar{N_i^c}N_j+\paren{y_{Nij}+y_{N2,ij}{H^\dagger H\over \Lambda_{N2}^2}}\bar{N}_i H L_j+h.c.
}
We evaluate the order of the resultant asymmetry obtained by the decay of the SM neutrino again.
The set of the thermal initial conditions of the Boltzmann equations is
\al{
&
N_X(z_\text{ini})
=
{3\over4},
&
N_{B-L}(z_\text{ini})
=
0,
}
In this case, the decay rate is\footnote{The last factor appears because of the phase space integral.}
\al{
\Gamma_X\fn{z}=  \vev{\frac{1}{\gamma}}\frac{M_{\nu}}{8\pi}
\left( 
\red{g_2^2+}
|Y_2|^2 
+\frac{\vev{h}^2}{2\Lambda^2} 
+\paren{\vev{h}^3\over2\sqrt{2}\Lambda_3^3}^2
\right)
\sqrt{1-\paren{M_W^2\over M_\nu^2}}
,
}
and the functions which appear in Boltzmann equation are roughly given by
\al{
\epsilon_i&\simeq
\sum_{j} 
{\text{Im}\sqbr{\paren{Y_2 Y_2^\dagger}^2}_{ij}\over8\pi \paren{Y_2 Y_2^\dagger}_{ij}},&
\vev{\sigma_\text{ann}v}&\simeq \alpha_2^2{1\over\text{Max}(M_\nu^2,T^2)},&
\nn
\vev{\sigma_L v}&\simeq \left({|Y_2|^2\over4\pi}\right)^2{1\over\text{Max}(M_\nu^2,T^2)},&
\text{Br}&\simeq{\red{1} \over\red{g_2^2\over |Y_2|^2}+1+{\vev{h}^2\over2|Y_2|^2\Lambda^2}+\paren{\vev{h}^3\over2\sqrt{2}|Y_2|\Lambda_3^3}^2},&
}
where we define the effective charged lepton Yukawa coupling $Y_{2ij}:=Y_{ij}+y_{2,ij}\vev{h}^2/(2\Lambda_2^2)$, and the neutrino mass $M_{\nu i}:=\text{diag}\paren{\lambda_{ij}\vev{h}^2/\Lambda+y_{3,ij}\vev{h}^4/(2\Lambda_3^3)}_i$.
We further assume that the components of $M_{\nu i}$ and $Y_{2ij}$ are the same order of magnitude, respectively, and we denote $M_{\nu i}=M_\nu, Y_{2ij}=Y_2$ for simplicity.
We focus on the asymmetry generated by the lightest neutrino in the false vacuum.
In Fig.~\ref{Fig:charged_Higgs}, we show the result assuming that  $CP$ phase is of the order of one, i.e. $e^{i\delta}\sim 1$. 
We use the following parameter set to draw the plot:
\al{\Lambda&=6\times10^{14}\, \GeV, \,\,&
\Lambda_2&=6\times10^{13}\, \GeV, \, \,&
\Lambda_3&=3\times10^{13}\, \GeV,\,\,&
\nn
\label{Eq:parameter set}
\vev{h}&=2\times10^{13}\, \GeV,\,\,&
y_{2ij}&\simeq1,\,\,&
\red{y_{3,ij}}&\red{=}\red{1}.&
}
We can see that the BAU is reproduced in this extension.
%
%

Notice that, unlike the minimal model, we obtain
\al{\label{Eq:extended model analytic}
\epsilon_i \text{Br}
\sim
\red{3\times10^{-6}},
}
in the extended model with the parameters in Eq.~\eqref{Eq:parameter set}.
This value is much larger than that in Eq.~\eqref{Eq:minimal model analytic}. This is one reason why we can obtain the realistic baryon asymmetry in the extended model.
Numerically, the resultant asymmetry becomes smaller than Eq.~\eqref{Eq:extended model analytic} due to the washout effect of inverse decay process, as in the standard baryogenesis scenario by the decay of heavy particle.

%
%
%


\section{Thermal history}\label{Thermal history}
In this section, we discuss the thermal history of the universe.
We introduce a new scalar $S$ to make \red{the} Higgs field stay at false vacuum in the early universe, where $S$ is singlet under the SM gauge group.\footnote{In the mechanism we have proposed, it is important that the Higgs field obtains a large expectation value at higher temperature. To realize this situation, we introduce this singlet-scalar field $S$. If one can realize this situation by other ways, we do not have to introduce it. But, naively, introducing the singlet-scalar field is a simplest way.}

First, we explain the zero temperature scalar potential of the extended model with $S$ and the thermal correction to it.
Then, we discuss how the Higgs field is in false vacuum in the early universe.

\subsection{Zero temperature Higgs potential}
The tree level scalar potential is given by
\al{\label{Eq:tree_pot}
V _\text{tree}\fn{h,S}=
-\kappa{m_S^2\over4\lambda_S} h^2+{1\over4}\lambda\, h^4 + \kappa h^2 S^2 - {1\over2}m_S^2 S^2 + \lambda_S S^4
,
}
where $S$ is the new singlet scalar field.
%
%
We consider the region where all couplings take $\mathcal{O}(0.1\text{--}1)$ value.
Although $\lambda$ becomes small or negative at high scale in the SM (see e.g. Ref.~\cite{Hamada:2012bp}), now the running of $\lambda$ is modified, $\lambda$ can take $\mathcal{O}(0.1\text{--}1)$ value since some scalar fields are added.

We note that the one-loop Coleman--Weinberg potential can be \red{safely} neglected because of $\mathcal{O}(0.1\text{--}1)$ couplings, and therefore we do not include it for simplicity.

The potential~\eqref{Eq:tree_pot} has an absolute minimum at\footnote{
The potential~\eqref{Eq:tree_pot} has a minimum at $\vev{h}=\sqrt{\frac{\kappa m_S^2}{2\lambda \lambda_S}}=\sqrt{\frac{2\kappa}{\lambda}}v_S$, $\vev{S}=0$.
This minimum does not becomes the absolute minimum but the local one for the parameter space we consider here.
}
\al{\label{Eq:true_minimum}
\vev{h}&=0,& 
\vev{S}&={1\over2}\sqrt{\frac{m_S^2}{\lambda_S}}\equiv v_S.&
}
The quadratic term of the SM Higgs is added in order to make the Higgs massless in this vacuum.

\subsection{Thermal potential}
We follow the Ref.~\cite{Carrington:1991hz} and show the thermal potentials.
The thermal potentials are evaluated at the one-loop level where the loop effects of the massive Higgs boson, $W$, $Z$ boson, the top quark and the scalar $S$ are included.
For the gauge fields, we employ the Landau gauge where the ghost fields are massless and do not have the $h$ field dependence.
The NG bosons $\chi_i$ in the Higgs doublet field \eqref{higgsdoublet} are neglected since their effects are small.

As the thermal effects, there are two components, namely $V_{\rm FT}\fn{h,T}$ and $V_{\rm ring}\fn{h,T}$.\footnote{The derivation of these functions is shown in appendix~\ref{The effective potential at finite temperature}.}
The main contribution of thermal effects comes from $V_{\rm FT}\fn{h,T}$, which is
\al{
V_{\rm FT}\fn{h,T}
&=\frac{T^4}{2\pi^2}\bigg[
J_B\fn{\tilde m_S^2/T^2}
+J_B\fn{ {\tilde m}_h^2 /T^2}
+6J_B\fn{ {m}_W^2 /T^2} 
+ 3J_B\fn{{m}_Z^2 /T^2} 
-12J_F\fn{ {m}_t^2/T^2}\bigg],
\label{VT}
}
where the mass for each particle is given by
\al{
{m}_W^2 &= \frac{g_2^2}{4}h^2,&
{m}_Z^2&= \frac{g_2^2+g_Y^2}{4} h^2,&
{m}_t^2 &= \frac{y_t^2}{2}h^2,&\nn
{\tilde m}_h^2 &=3\lambda h^2-\kappa\frac{m_S^2}{2\lambda_S}+2\kappa S^2,&
{\tilde m}_S^2 &=12\lambda_S S^2 +2\kappa h^2 -m_S^2;&
\label{mh2}
}
the thermal functions are defined as
\al{
J_B \fn{r^2} &=\int_0^\infty dx\, x^2 \ln
\left(1-e^{-\sqrt{x^2+r^2} } \right),&
J_F\fn{r^2} &=\int_0^\infty dx\, x^2 
\ln\left(1+e^{-\sqrt{x^2+r^2} } \right).&
\label{thermal functions exact}
}
Remember here that the coupling constants $g_2$, $g_Y$ and $y_t$ are ${\rm SU}(2)_L$, ${\rm U}(1)_Y$ and top-Yukawa coupling constants, respectively.
Since one cannot analytically and exactly evaluate these functions, the approximated expressions are made.\footnote{
The high temperature expansion is often used. However, they are not useful for the case where we see the large field value of $h$.
Therefore, the fitting functions \eqref{fitting function} are also employed~\cite{Funakubo:2009eg}.
See appendix~\ref{The effective potential at finite temperature} for details.
}

There are contributions to the ring diagrams (or the daisy diagrams) from the Higgs boson and the gauge boson:
\al{
V_{\rm ring}\fn{h,T}
&= - \frac{T}{12\pi} \bigg[ \left( \tilde m^2_h +\Pi_h\fn{T} \right)^{3/2} - \tilde m^3_h \bigg] 
- \frac{T}{12\pi} \bigg[ \left(\tilde m^2_S +\Pi_h\fn{T} \right)^{3/2} - \tilde m^3_S \bigg] 
\nn
&\quad -\frac{T}{12\pi}\bigg[
2 a_g^{3/2}+\frac{1}{2\sqrt{2}}\left(a_g+c_g-[(a_g-c_g)^2+4 b_g^2]^{1/2}\right)^{3/2}\nn
&\qquad 
+\frac{1}{2\sqrt{2}}\left(a_g+c_g+[(a_g-c_g)^2+4 b_g^2]^{1/2}\right)^{3/2}-\frac{1}{4}[g_2^2 h^2]^{3/2}
-\frac{1}{8}[(g_2^2+g_Y^2) h^2]^{3/2}\bigg],
}
where the first and second terms correspond to the contribution from the Higgs and the scalar $S$;\footnote{Combining the ring contribution of the Higgs boson and the first term of Eq.~\eqref{VT}, we can write
\al{
\frac{T^4}{2\pi^2}J_B\fn{ {\tilde m}_h^2 /T^2} - \frac{T}{12\pi} \bigg[ \left( \tilde m^2_h +\Pi_h\fn{T} \right)^{3/2} - \tilde m^3_h \bigg] 
= \frac{T^4}{2\pi^2}J_B\fn{ {\tilde m}_h^2\fn{T} /T^2},
}
where $\tilde m^2_h\fn{T} = \tilde m_h^2 + \Pi_h\fn{T}$ is the Debye mass of the Higgs boson.
In the same manner,  the thermal effects for the scalar $S$ also can be written as the same form.
}
the thermal masses of the Higgs and scalar $S$ are
\al{
\Pi_h\fn{T} &= \frac{T^2}{12}\left(\frac{9}{4}g_2^2+
\frac{3}{4}g_Y^2+3 y_t^2+6\lambda + \red{2}\kappa\right),
\label{thermal mass of higgs}
\\
\Pi_S\fn{T} &= T^2\left(\frac{\lambda_S}{4} + \frac{\red{2}\kappa}{3} \right);
\label{thermal mass of s}
}
and we have defined 
\al{
a_g &=\frac{1}{4}g_2^2 h^2+\frac{11}{6}g_2^2 T^2,&
b_g &= -\frac{1}{4}g_2 g_Y h^2,&
c_g &= \frac{1}{4}g_Y^2 h^2+\frac{11}{6}g_Y^2 T^2.&
\label{definitions of ring}
}

To summarize,  in order to trace the thermal history of the Higgs potential in the SM, we analyze the effective potential
\al{
V_{\rm eff}\fn{h,S,T} &=V_{\rm tree}\fn{h,S}+
V_{\rm FT}\fn{h,S,T}+V_{\rm ring}\fn{h,S,T},
\label{VEFF}
}
where $V_{\rm tree}\fn{h,S}$ is given in Eq.~\eqref{Eq:tree_pot}.
In next subsection, we investigate the phase transition of Higgs field by using this potential.

\subsection{Thermal history}
In the early universe, due to the finite temperature effect, $S$ and $H$ \red{do} not have the vacuum expectation value(VEV).\footnote{Our thermal scenario is similar to Ref.~\cite{Jinno:2015doa} where the gravitational wave from electroweak phase transition at the high scale is discussed.}
They develop \red{their respective VEVs} at the temperature when the thermal mass term becomes comparable with \red{their} negative mass term.
By utilizing the high temperature expansion \eqref{JBexpansion} and \eqref{JFexpansion}, we estimate the critical temperatures which are given as the vanishing curvature of $V_\text{eff}\fn{h,S,T}$ at the origin $(h,S)=(0,0)$, namely
\al{
\frac{\p^2 V_\text{eff}\fn{h,S,T_S}}{\p S^2}\bigg|_{h=0,S=0}&=0,&
\frac{\p^2 V_\text{eff}\fn{h,S,T_h}}{\p h^2}\bigg|_{h=0,S=0}&=0.&
}
Solving these equations for $T$, we find\footnote{
\red{
Here the $V_\text{ring}$ contribution is neglected.
This should still provide an approximate estimation of the phase transition temperature.
}
}
\al{
T_S&=
\frac{2\sqrt{3}\sqrt{2v_s^2\lambda_S}}{\sqrt{\kappa+6\lambda_S}}
,&
T_h&=
\frac{4\sqrt{6}\sqrt{v_s^2\kappa }}{\sqrt{9g_2^2+3g_Y^2+12y_t^2+8\kappa+12\lambda}}
.&
}
Here $T_S$ and $T_h$ denote the critical temperatures of the phase transition of $S$ and $h$, respectively.
Our scenario is as follows.
The phase transition of Higgs field happens at $T=T_h$. 
At this time, $S$ and $h$ are in the false vacuum, $\vev{S}=0, \vev{h}=\sqrt{2\kappa\over\lambda}v_S$, and the lepton number is created by the decay of heavy neutrinos.
After that, at $T=T_S$, $S$ develops VEV $\vev{S}=v_S$, and then $\vev{h}$ comes back to the true vacuum Eq.~\eqref{Eq:true_minimum}.

In order to work with our scenario, we require
\al{
T_S
< 
T_h 
.
}
Moreover, $S$ must have a negative mass at $\vev{S}=0, \vev{h}=\sqrt{2\kappa\over\lambda}v_S$, namely $\tilde m_S <0$, which yields
\al{\label{Eq:second}
\lambda_S>{\kappa^2\over\lambda}.
}

As an example of successful parameters, we take $\kappa=0.7, \lambda_S\simeq1.5, \lambda=0.4$ and $\vev{h}=2\times10^{13}\GeV$.
$T_h$ and $T_S$ become
\al{&
T_S
\simeq 1.9 v_S, 
&
T_h
\simeq 2.0 v_S
,
}
and Eq.~\eqref{Eq:second} is satisfied.
Here $g_Y=g_2=y_t=0.5$ is used.

Therefore, by solving the Boltzmann equations with
\al{
&
z_\text{ini}
=
{M_{\nu}\over T_h}
,
&
z_\text{final}
=
{M_{\nu}\over T_S}
,
}
we can calculate the asymmetry.
For example, we obtain\footnote{Here we take the thermal initial condition, $N_X=3/4, N_{B-L}=0$.}
\al{
N_{B-L}\simeq 7.0\times 10^{-7},
}
with the parameter set Eq.~\eqref{Eq:parameter set}.
%
%
Here we have taken into account the washout factor~\cite{Hamada:2015xva} in the symmetric phase,
\al{
\exp[-T_S/2\times10^{13}\GeV].
}
%
This implies that we can realize the observed value, $N_{B-L,\text{obs}}=4.8\times 10^{-8}$, by slightly changing the value of $CP$ phase.
\red{
We notice that a numerical study is necessary to establish which values of the couplings return an acceptable pattern of symmetry breaking, as currently approximate estimates are provided in the paper.
}

Finally, let us briefly discuss the validity of the effective Lagrangian Eq.~\eqref{dimfiveop}.
The temperature of the phase transition is 
\al{
T_h\simeq \sqrt{2\lambda\over\kappa}\vev{h}\simeq 2\times10^{13}\GeV,
} 
while the cutoff scale in Eq.~\eqref{dimfiveop} is Eq.~\eqref{cutoff}.
It can be seen that, as long as $m_\nu\lesssim 0.1\eV$, $T_h$ is much smaller than $\Lambda$ in Eq.~\eqref{Eq:parameter set}.  Although $T_h$ is close to $\Lambda_2$ and $\Lambda_3$, it is still below these cutoffs. Hence, the effective Lagrangian would be valid in this region.

\section{Summary and discussion}\label{Summary and discussion}
We have considered the possibility of baryogenesis in a false vacuum where the Higgs field develops a large field value compared with the electroweak scale.
Since all the SM particles receive mass from the coupling with the Higgs boson, the large field value of the Higgs field means that they are super-heavy. 
We have estimated the asymmetry produced by the decay of the heavy left-handed neutrino.
It has turned out that the decay of the neutrino can not realize the observed baryon asymmetry.
If the new higher-dimensional operators are introduced, the decay of the neutrino can provide sufficient asymmetry.

We have also presented the thermal history where the Higgs field develops a large field value in the early universe.
It has been found that, by adding the singlet scalar $S$, our scenario safely works.

Finally, we briefly mention the possibility of the high scale electroweak baryogenesis.
So far, we have pursued the possibility that the baryon asymmetry is created by the heavy particle while the lepton number violation is given by Majorana mass term of the left-handed neutrino.
However, if the coupling $\lambda$ is small, the electroweak phase transition at high scale becomes of first-order.
Since our extended model has many $CP$ phases, there is a possibility to generate the $B+L$ asymmetry.
If the $L$ asymmetry is washed out in the false vacuum, the net $B$ asymmetry survives.
The condition of the $L$ wash out would be roughly given by
\al{
\paren{\vev{h}\over\Lambda}^2{1\over8\pi}M_W\gtrsim\sqrt{10}{T^2\over M_P}.
}
By putting $T\simeq\vev{h}, M_W\simeq\vev{h}$, we obtain
\al{
\vev{h}
\gtrsim1\times 10^{13}\GeV.
}
Hence, we have a possibility to create the baryon asymmetry by the electroweak baryogenesis in addition to the decay of heavy particle.

\subsection*{Acknowledgement}
%
YH is supported by Japan Society for the Promotion of Science (JSPS) Fellowships for Young Scientists.
We thank Hikaru Kawai for important discussion of the phase diagram of the standard model.
We also thank Ryusuke Jinno and Masahiro Takimoto for useful comments on the Higgs thermal potential.

\begin{appendix}
\section{The effective potential at finite temperature}\label{The effective potential at finite temperature}
In this appendix, following Ref.~\cite{Carrington:1991hz}, we show the derivation of the thermal effects on the Higgs potential in the SM.
We consider the one-loop contribution of a particle with the mass $m\fn{h}$ to the potential, which typically has the following form:
\al{
V_{\rm 1loop}\fn{h,T}=\pm \frac{1}{2} {\rm Tr}\ln\fn{ {k}^2 + m^2\fn{h}},
}
where ``${\rm Tr}$" denotes the functional trace; $k$ is the Euclidean momentum; the boson (fermion) loop case has overall positive (negative) sign. 
For a particle with one degree of freedom, the potential is
\al{
V_{\rm 1loop}\fn{h,T}=\pm \frac{1}{2}\int \frac{d^4 k}{(2\pi)^4}\, \ln\fn{ {k}^2 + m^2\fn{h}}.
\label{loop potential}
}
At finite temperature, the time direction of momentum is discretized and its loop integral changes to the Matsubara summation:
\al{
\int \frac{dk_0}{2\pi} \int \frac{d^3 k}{(2\pi)^3}\, f\fn{k_0,\vec k}=T \sum_{n=-\infty}^\infty \int \frac{d^3 k}{(2\pi)^3}\, f\fn{\omega_n, \vec k},
}
with the Matsubara frequency,
\al{
\omega_n = 
\begin{cases}
2 n\pi T & \text{for boson,}\\
(2n+1)\pi T & \text{for fermion.}
\end{cases}
}
Therefore, the Eq.~\eqref{loop potential} can be calculated as
\al{
V_{\rm 1loop}\fn{h,T}& = \pm \frac{T}{2} \int \frac{d^3 k}{(2\pi)^{3}} \left[ \beta \omega +2\ln (1\mp e^{-\beta \omega}) \right]\nn
&= \pm\frac{1}{2}\int \frac{d^3 k}{(2\pi)^{3}}\, \omega \pm T \int \frac{d^3 k}{(2\pi)^{3}}\ln (1\mp e^{-\beta \omega}),
} 
where $\omega=\sqrt{\vec k^2 + m^2}$; the sign $(+)$ and $(-)$ in the logarithm apply to fermions and bosons, respectively. 
The first term does not depend on temperature and is rewritten as
\al{
V_{\rm CW}\fn{h} &\equiv \pm \frac{1}{2}\int \frac{d^3 k}{(2\pi)^{3}}\, \omega \nn
&= \pm \frac{1}{2}\int \frac{d^4 k}{(2\pi)^4}\, \ln\fn{k^2 + m^2 }.
}
This is one-loop contribution at vanishing temperature, i.e. the Coleman--Weinberg potential.
The second term is the thermal potential at one-loop level and becomes
\al{
V_{\rm FT}\fn{h,T}\equiv
\pm \frac{T}{2\pi^2} \int dk\, k^2\ln (1\pm e^{-\beta \omega})
=\pm \frac{T^4}{2\pi^2} J_{B(F)}\fn{r^2}, 
}
where the thermal functions for boson and fermion are defined as 
\al{
J_B \fn{r^2} &=\int_0^\infty dx\, x^2 \ln
\left(1-e^{-\sqrt{x^2+r^2} } \right),&
J_F\fn{r^2} &=\int_0^\infty dx\, x^2 
\ln\left(1+e^{-\sqrt{x^2+r^2} } \right),&
\label{thermalfunctionexact}
}
with $x\equiv |\vec k|/T$ and $r\equiv m\fn{h}/T$.
Note that in general case, the operator $k^2 + m^2\fn{h}$ is not diagonal, i.e. $k^2 \delta_{ij}+m_{ij}^2\fn{h}$.
Therefore, the mass matrix $m_{ij}^2\fn{h}$ has to be diagonalized.

In the SM case, taking account of the degrees of freedom of particles, the thermal potential is given by
\al{
V_{\rm FT}\fn{h,T}= \sum_{i=W,\, Z,\, h}n_i\frac{T^4}{2\pi^2} J_{B}\fn{(m^B_i\fn{h}/T)^2}
-\sum_{i=t}n_i\frac{T^4}{2\pi^2} J_{F}\fn{(m^F_i\fn{h}/T)^2},
}
where $n_W=6$, $n_Z=3$, $n_t=12$ and $n_h=1$.

\begin{figure}
\begin{center}
\hfill
\includegraphics[width=.6\textwidth]{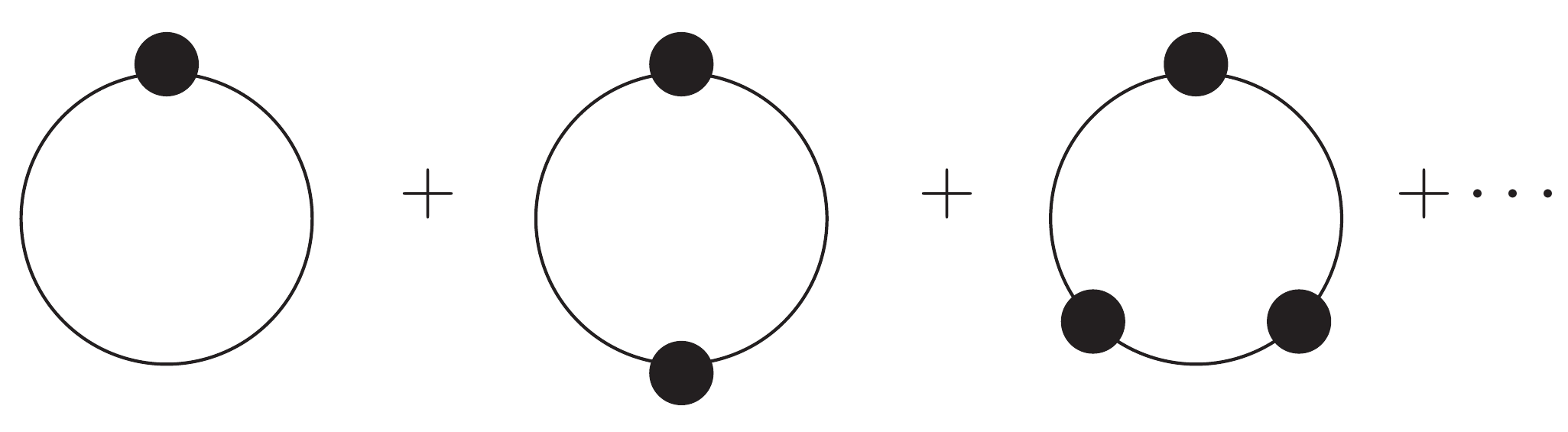}
\hfill\mbox{}
\end{center}
\caption{The ring diagrams. The black circle denotes the propagator with loop corrections.}
\label{Fig:ringdiagrams}
\end{figure}
\begin{figure}
\begin{center}
\hfill
\includegraphics[width=.7\textwidth
]{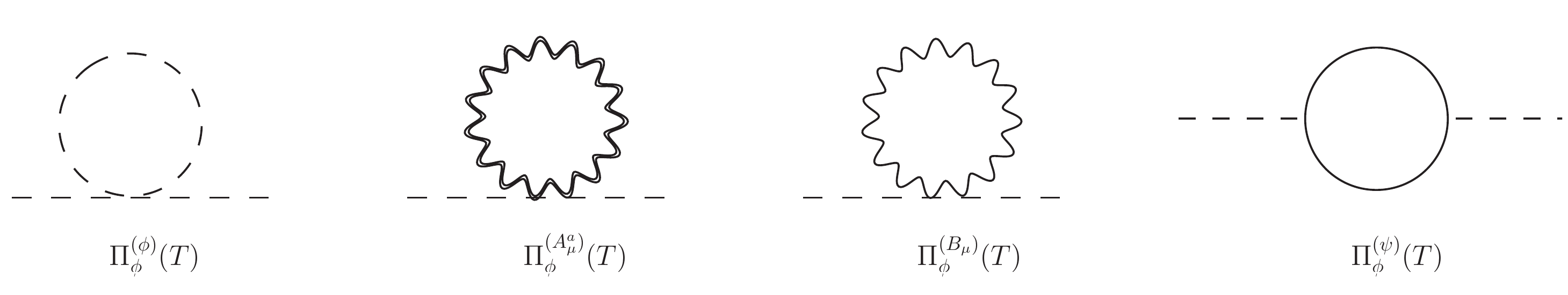}
\hfill\mbox{}
\end{center}
\caption{The two-point functions of Higgs field at one-loop level which yield the thermal masses. The dashed, single wave, double wave and solid lines denote the Higgs, $\text{U}(1)_Y$ gauge, $\text{SU}(2)_L$ and top-quark, respectively.}
\label{Fig:thermalmasshiggs}
\end{figure}
\begin{figure}
\begin{center}
\hfill
\includegraphics[width=.7\textwidth,
]{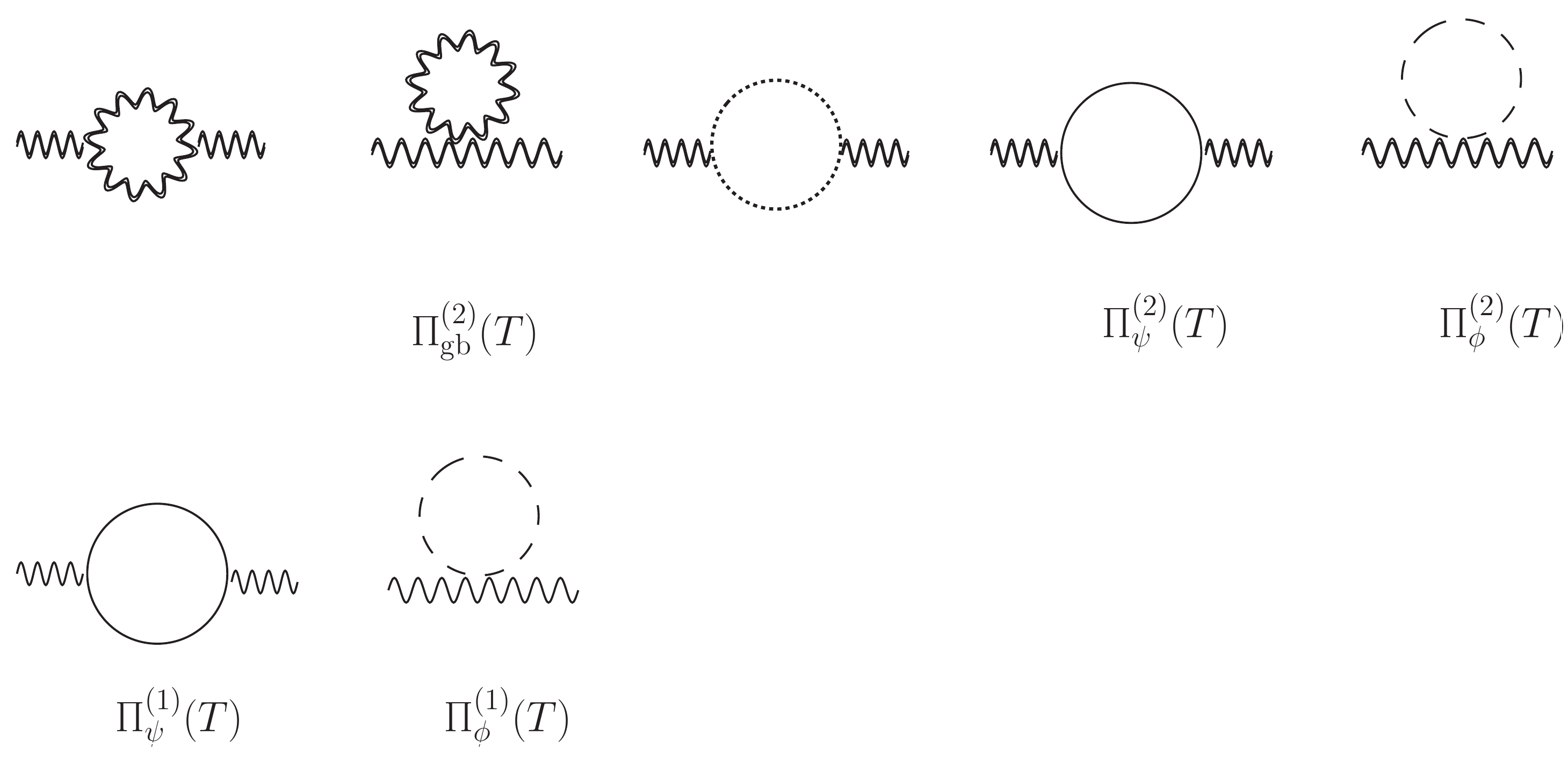}
\hfill\mbox{}
\end{center}
\caption{The two-point functions of $\text{SU}(2)$ and $\text{U}(1)$ gauge fields at one-loop level which yield the thermal masses. The dashed, single wave, double wave, solid and dot lines denote the Higgs, $\text{U}(1)_Y$ gauge, $\text{SU}(2)_L$, top-quark and ghost, respectively.}
\label{Fig:thermalmassgauge}
\end{figure}
Next, we consider the ring (or daisy) contributions shown in Fig.~\ref{Fig:ringdiagrams}, which are the next-higher-order corrections and are related to the infrared divergence; see e.g.~\cite{Kapusta:2006pm} for a detailed discussion.
The ring contribution for the Higgs field is given by
\al{
V_\text{ring}^\text{Higgs}\fn{h,T}
&\equiv -\frac{1}{2} T \sum_{n=-\infty}^{\infty} \int \frac{d^3 k}{(2\pi)^3}
\sum_{l=1}^\infty \frac{1}{l} 
\left( -\frac{1}{\omega_n^2 + {\vec k}^2 + m_h^2\fn{h}} \Pi_h\fn{T} \right)^l\nn
&=-\frac{T}{12\pi} \text{Tr}\, \left\{ [m_h^2\fn{h}+\Pi_h\fn{T}]^{3/2} -m_h^3\fn{h} \right\},
}
where the thermal mass comes from the diagrams in the limit $m\fn{h}/T \ll 1$ shown in Fig.~\ref{Fig:thermalmasshiggs} and becomes
\al{
\Pi_h\fn{T} =
\Pi^{(A_\mu^a)}_\phi\fn{T} + \Pi^{(B_\mu)}_\phi\fn{T} + \Pi^{(\psi)}_\phi\fn{T}  + \Pi^{(\phi)}_\phi\fn{T} 
= \frac{T^2}{12}\left(\frac{9}{4}g_2^2+\frac{3}{4}g_Y^2+3 y_t^2+6\lambda\right),
}
with
\al{
\Pi^{(A_\mu^a)}_\phi\fn{T} &= \frac{3}{16}g_2^2T^2,&
\Pi^{(B_\mu)}_\phi\fn{T} &= \frac{1}{16}g_Y^2T^2,&
\Pi^{(\psi)}_\phi\fn{T} &= \frac{1}{4}y_t^2T^2,&
\Pi^{(\phi)}_\phi\fn{T} &= \frac{1}{2}\lambda T^2.&
}
Note that these contributions are evaluated by setting the external momentum to zero since we are interested in the infrared limit.

In a similar manner, one can obtain the ring contributions from the gauge bosons, which becomes
\al{
V_\text{ring}^\text{gb}\fn{h,T} \equiv -\frac{T}{12\pi} \text{Tr}\, \left\{ [M^2\fn{h}+\Pi_{00}\fn{T}]^{3/2} -M^3\fn{h} \right\}.
}
Here the mass matrices in the original gauge field basis $(A_\mu^i, B_\mu)$ are
\al{
M^2\fn{h}&=\pmat{
g_2^2h^2/4 & 0 & 0 & 0\\
0 & g_2^2h^2/4 & 0 & 0\\
0 & 0 & g_2^2h^2/4 & -g_Yg_2h^2/4 \\
0 & 0 & -g_Yg_2h^2/4 & g_2^2h^2/4
},\\
\Pi_{00}\fn{T}&=\pmat{
\Pi^{(2)}_{00}\fn{T} & 0 & 0 & 0\\
0 & \Pi^{(2)}_{00}\fn{T}  & 0 & 0\\
0 & 0 & \Pi^{(2)}_{00}\fn{T}  & 0 \\
0 & 0 & 0 &\Pi^{(1)}_{00}\fn{T}
},\label{thermal mass matrix}
}
where $\Pi_{00}\fn{T}$ is the $(00)$ component of  the polarization tensor in the infrared limit, namely $\Pi_{\mu\nu}\fn{p=0,T}$ and 
\al{
\Pi^{(1)}_{00}\fn{T} 
&= \Pi^{(1)}_{\phi}\fn{T} + \Pi^{(1)}_\psi \fn{T} = \frac{11}{6}g_Y^2T^2, \\
\Pi^{(2)}_{00}\fn{T}
&= \Pi^{(2)}_\text{gb}\fn{T} + \Pi^{(2)}_{\phi}\fn{T} +\Pi^{(2)}_\psi \fn{T} = \frac{11}{6}g_2^2T^2,
}
with
\al{
\Pi^{(1)}_\phi \fn{T} &=\frac{1}{6}g_Y^2 T^2, 
& \Pi^{(1)}_\psi \fn{T} &=\frac{5}{3}g_Y^2 T^2, &\\
\Pi^{(2)}_\text{gb} \fn{T} &=\frac{2}{3}g_2^2 T^2, &
\Pi^{(2)}_\phi \fn{T} &=\frac{1}{6}g_2^2 T^2, &
\Pi^{(2)}_\psi \fn{T} &= g_2^2 T^2. &
}
These thermal masses are obtained by calculating the two-point functions of $\text{SU}(2)$ and $\text{U}(1)$ gauge fields shown in Fig.~\ref{Fig:thermalmassgauge}.
Evaluating the eigenvalues of $M^2\fn{h}+\Pi_{00}\fn{T}$ and $M^2\fn{h}$ to the three-half power, and then taking trace of them, we have
\al{
V_\text{ring}^\text{gb}\fn{h,T}&=
 -\frac{T}{12\pi}\bigg[
2 a_g^{3/2}+\frac{1}{2\sqrt{2}}\left(a_g+c_g-[(a_g-c_g)^2+4 b_g^2]^{1/2}\right)^{3/2}\nn
&\quad 
+\frac{1}{2\sqrt{2}}\left(a_g+c_g+[(a_g-c_g)^2+4 b_g^2]^{1/2}\right)^{3/2}-\frac{1}{4}[g_2^2 h^2]^{3/2}
-\frac{1}{8}[(g_2^2+g_Y^2) h^2]^{3/2}\bigg],
}
where $a_g$, $b_g$ and $c_g$ are given in Eq.~\eqref{definitions of ring}.

Note that we have worked in the Landau gauge to evaluate the contributions from the gauge bosons.
Although the thermal mass matrix \eqref{thermal mass matrix} can be diagonal only in the limit $m_W\fn{h}/T\ll 1$ and $m_Z\fn{h}/T\ll 1$, the ring contribution is still valid for the larger mass, thus the larger field value than temperature.
This is because the ring contribution vanishes for the larger mass.

\begin{figure}
\begin{center}
\hfill
\includegraphics[width=.4\textwidth]{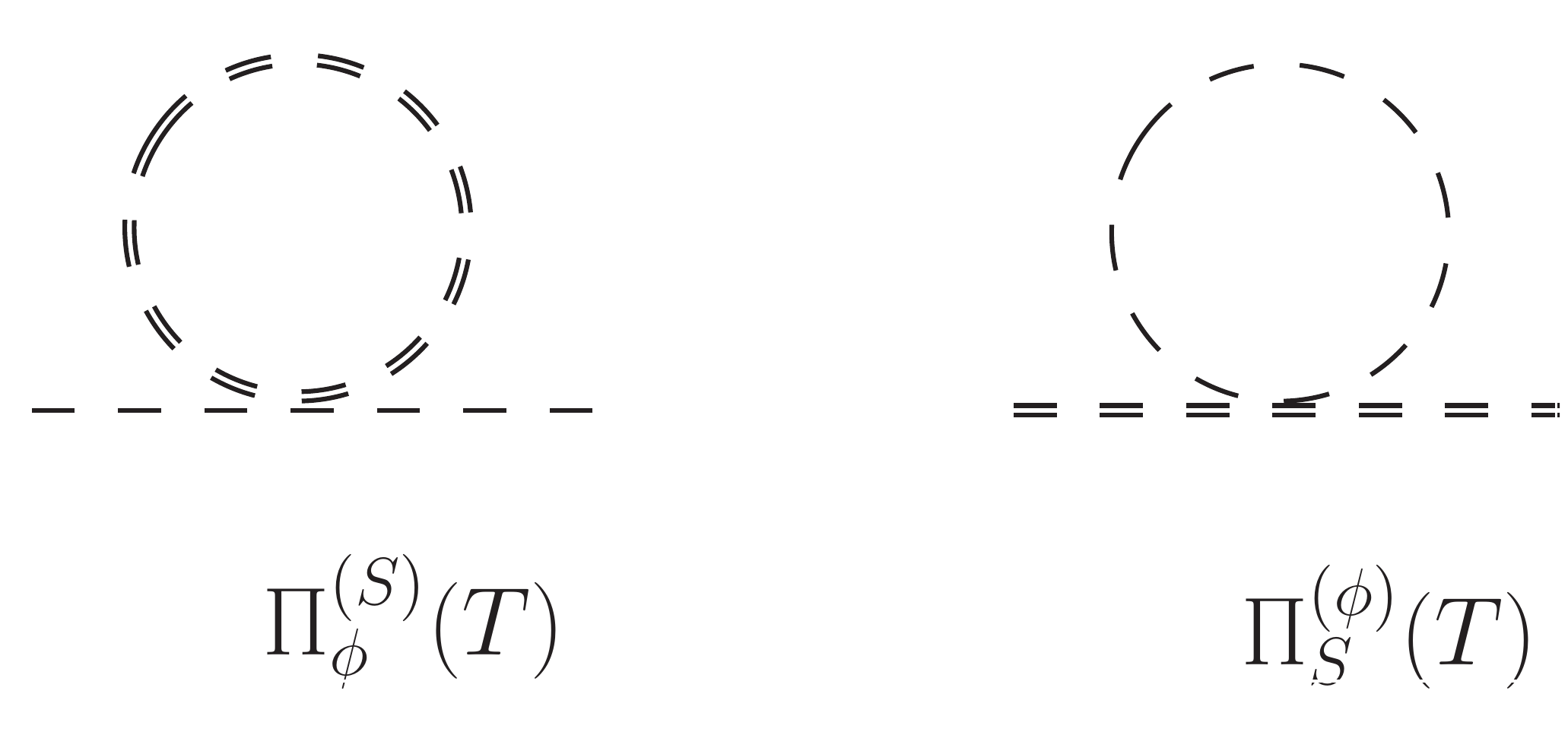}
\hfill\mbox{}
\end{center}
\caption{The diagrams which contribute to the thermal masses of the Higgs field and the new scalar one. The double dashed line denotes the new scalar field $S$.}
\label{Fig:thermalmassnewandhiggs}
\end{figure}
In case where the scalar $S$ is introduced, the contribution from the diagram shown in Fig.~\ref{Fig:thermalmassnewandhiggs} is added, and then the thermal masses of the Higgs field and $S$ are given as Eq.~\eqref{thermal mass of higgs} and \eqref{thermal mass of s}, respectively.

\subsection{The thermal functions and their approximation}
The high temperature expansion is often applied to the thermal functions~\eqref{thermalfunctionexact}.
However, it is not useful for investigating the large field value $m\fn{\phi}/T \equiv r\geq 1$.
In this subsection we compare the exact forms of the the thermal functions~\eqref{thermalfunctionexact} numerically evaluated with their approximated forms and investigate the effectiveness of them.

The thermal functions~\eqref{thermalfunctionexact} with the high temperature expansion become
\al{
J_B \fn{r^2} &\simeq 
-\frac{\pi^4}{45}+\frac{\pi^2}{12}r^2
-\frac{\pi}{6}r^{3}-\frac{r^4}{32}
\left[\ln (r^2 /16\pi^2)+2\gamma_E-\frac{3}{2}   \right],
\label{JBexpansion}
\\
J_F\fn{r^2} &\simeq 
\frac{7\pi^4}{360}-\frac{\pi^2}{24}r^2
-\frac{r^4}{32}\left[\ln (r^2 /\pi^2)+2\gamma_E-\frac{3}{2}
   \right],
   \label{JFexpansion}
}
where $\gamma_E\approx 0.57721$ is the Euler's gamma. 
Besides, it is known that the thermal functions can be fitted by the following functions~\cite{Funakubo:2009eg}:
\al{\label{fitting function}
J_{B(F)}\fn{r^2} &= e^{-r} \sum_{n=0}^{N_{B(F)}} c^{B(F)}_nr^n,
}
where $N_{B(F)}$ and $c^{B(F)}_n$  are the truncation order of the series and the fitting coefficients, respectively.
For the fixed truncation order $N_{F(B)}$, we find the coefficients $c^{F(B)}_n$ by fitting the exact results numerically evaluated.

\begin{figure}
\begin{center}
\includegraphics[width=10cm]{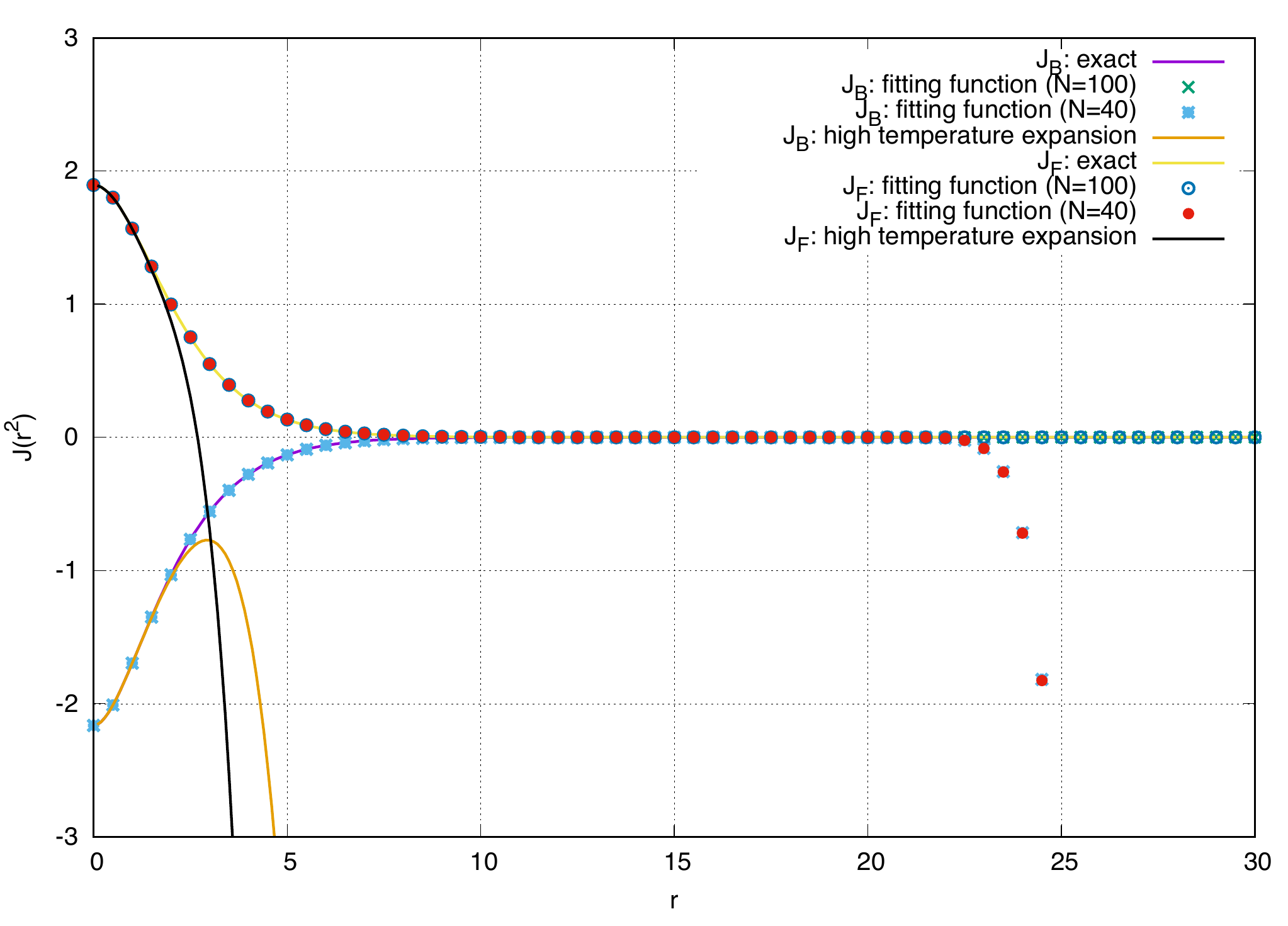}
\caption{Comparison of the thermal functions between different approximations.}
\label{thermalfunctionsplot}
\end{center}
\end{figure}

We show the comparisons between results of exact form Eq.~\eqref{thermalfunctionexact} and the approximated forms Eq.~\eqref{JBexpansion}, \eqref{JFexpansion} and \eqref{fitting function} in Fig.~\ref{thermalfunctionsplot}.
We see that the high temperature expansions are actually valid for $r\leq 2$ and the fitting functions with $N_{B(F)}=40$ break down for $r>22$.
The fitting functions with $N_{B(F)}=100$ are valid for large value of $r$.
Therefore, it is useful for evaluating a potential of large field values since $r=m\fn{\phi}/T$ and the mass $m\fn{\phi}$ is proportional to the value of the field $\phi$.
\end{appendix}

\bibliographystyle{TitleAndArxiv}
\bibliography{refs}
\end{document}